\documentstyle[12pt]{article}
\def\gz{\mbox{${\sf Z \!\!\!\!\: Z}$}}
\def\rz{\mbox{${\sf I \! R}$}}
\def\d{{\rm d}}
\def\e{{\rm e}}
\def\i{\ifmmode{\rm i}\else\char"10\fi}
\def\I{{\rm I}}
\def\lesssim{\raisebox{-1mm}{\shortstack[c]{$<$\\[-1mm] $\sim$}}}
\def\gtrsim{\raisebox{-1mm}{\shortstack[c]{$>$\\[-1mm] $\sim$}}}
\begin{document}
{}~\\[5mm]
\noindent{\Large\bf Exact thermostatic results for
the $n$-vector model\\[2mm] on the harmonic chain}\\[7mm]
\noindent{\sl Georg Junker} and {\sl Hajo Leschke}\\[5mm]
Institut f\"ur Theoretische Physik, Universit\"at
Erlangen-N\"urnberg,\\ Staudtstr.\ 7, D-91058 Erlangen, Germany\\[10mm]

In this paper we report exact results on thermostatic properties of the
classical $n$-vector model on the harmonic chain. This system is characterized
by the Hamiltonian
\begin{equation}
H:=H_{0}\Bigl(\{p\},\{q\}\Bigr)+H_{1}\Bigl(\{q\},\{\vec{S}\}\Bigr)
\end{equation}
where
\begin{eqnarray}
H_{0}\Bigl(\{p\},\{q\}\Bigr)&:=&\frac{1}{2m}\sum_{j=1}^{N+1}p_{j}^{2}+
\frac{m}{2}\,\omega_{0} ^{2}\sum_{j=1}^{N}(q_{j}-q_{j+1})^{2},\\
H_{1}\Bigl(\{q\},\{\vec{S}\}\Bigr)&:=&-\sum_{j=1}^{N}
W(\ell +q_{j+1}-q_{j})\vec{S}_{j}\cdot\vec{S}_{j+1}.
\end{eqnarray}
Here $H_{0}$ is the Hamiltonian for the nearest-neighbor coupled harmonic
chain which consists of a one-dimensional lattice $\ell \gz$ with lattice
constant $\ell >0$ and a set of $N+1$ point particles of mass $m>0$ distributed
along the Euclidean line $\rz$ at positions $j\ell +q_{j}$, $j=1,2,\ldots,
N+1$.
The momentum of the $j$-th particle is denoted by $p_{j}$ and the spring
constant of this chain is $m\omega ^2_{0}$.
We assume now that each particle carries a set of internal rotational degrees
of freedom which we collectively represent by a classical spin, that is, by
an $n$-component Euclidean unit vector $\vec{S}_{j}\in\rz^{n}$.
The Hamiltonian $H_{1}$ then models the simplest rotational invariant
interaction between the spins of two nearest neighboured particles.
The interaction strength between two spins is described
by the real-valued even function $W:x\mapsto W(x)$ and depends on the actual
interparticle distance as indicated in (3). In accordance with the harmonic
approximation it is sufficient to consider only the first two terms in a
Taylor expansion of $W$,
\begin{equation}
W(\ell +q_{j+1}-q_{j})\approx J+(q_{j+1}-q_{j})\eta ,
\end{equation}
where $J:=W(\ell )$ and $\eta :=W'(\ell )$.

Within this approximation it is possible to decouple the vibrational and
rotational degrees of freedom by introducing shifted particle positions (cf.\
\cite{MS63,M85} for $n=1$)
\begin{equation}
x_{1}:=q_{1},~~~x_{j}:=q_{j}-\frac{\eta }{m\omega_{0} ^{2}}
\sum_{r=1}^{j-1}\vec{S}_{r}\cdot\vec{S}_{r+1}~,~~~j=2,3,\ldots ,N+1 .
\end{equation}
The result can be cast into the form
\begin{equation}
H\approx H_{0}\Bigl(\{p\},\{x\}\Bigr)+H_{\rm spin}\Bigl(\{\vec{S}\}\Bigr)
\end{equation}
where we have introduced the pure spin-chain Hamiltonian
\begin{equation}
H_{\rm spin}\Bigl(\{\vec{S}\}\Bigr):=
-\sum_{j=1}^{N}\left( J\vec{S}_{j}\cdot\vec{S}_{j+1}+
K(\vec{S}_{j}\cdot\vec{S}_{j+1})^{2}\right)
\end{equation}
with $K:=\eta ^{2}/(2m\omega_{0} ^{2})$.
Since the thermal properties of the harmonic chain are well known (see, for
example, \cite{AM75}), we will consider only those of  $H_{\rm spin}$.

Special cases of the Hamiltonian (7) have already been
discussed in the literature. For $K=0$ it corresponds to Stanley's
$n$-vector model in one dimension \cite{S69,S75}. For $J=0$ and $n\in\{2,3\}$
a discussion is due to Vuillermot and Romerio \cite{VR73}.
As for the Ising case ($n=1$), we remark that the biquadratic term in (7)
lowers the specific free energy of the Ising chain simply by the constant $K$.
However, even for $n=1$ this term is responsible for magnetostrictive effects
of the full system (6), as discussed by Mattis and Schultz \cite{MS63,M85}.
As an aside we mention that exactly known \cite{AKLT87} ground-state
properties of the quantum version of (7) for $n=3$,
$\hat{\vec{S}_{j}^2}=2\hbar^2\hat{\bf 1}$, $J<0$ and $K/|J|=-1/3$ are
discussed \cite{AKLT87} in relation with Haldane's conjecture.\\[3mm]

The basic thermostatic properties of the classical spin-chain Hamiltonian (7)
can be obtained from
the free energy per spin in the macroscopic limit $N\to\infty $:
\begin{equation}
F(\beta ):=-\frac{1}{\beta }\,\lim_{N\to\infty }\frac{1}{N+1}\,\ln
Z(\beta ),
\end{equation}
where the canonical partition function at temperature $1/\beta k$
($k$: Boltzmann's constant) for the finite chain and $n\geq 2$ may be defined
by the $(N+1)(n-1)$-dimensional integral
\begin{equation}
Z(\beta ):=\int\d \vec{S}_{1}\cdots
\int\d \vec{S}_{N+1}\,
\exp\left\{-\beta H_{\rm spin}\Bigl(\{\vec{S}\}\Bigr)\right\}.
\end{equation}
For convenience we are using open boundary conditions. Furthermore, each of
the above $\d\vec{S}$ stands for the usual surface measure on the
$(n-1)$-dimensional unit sphere in $\rz^{n}$.
We assume this measure to be normalized in the sense that $\int\d\vec{S}=1$
and recall its invariance under rotations
\begin{equation}
\int\d\vec{S}\, f(\vec{S})=
\int\d\vec{S}\, f(g\vec{S}).
\end{equation}
This relation is valid for any integrable complex-valued function $f$ and
all orthogonal $n\times n$ matrices $g\in SO(n)$. Of course, for $n=1$ the
integration $\int\d\vec{S}$ stands for the summation
$\frac{1}{2}\sum_{S=\pm 1}$.

For the evaluation of the partition function (9) we note that the Hamiltonian
(7) can be rewritten as
$H_{\rm
spin}\Bigl(\{\vec{S}\}\Bigr)=\sum_{j=1}^{N}V(\vec{S}_{j},\vec{S}_{j+1})$
where we have introduced the spin-pair interaction energy
\begin{equation}
V(\vec{S},\vec{S'}):=-J\vec{S}\cdot\vec{S}'-
K\left(\vec{S}\cdot\vec{S}'\right)^{2},
\end{equation}
which is $SO(n)$-invariant and exchange-invariant:
\begin{equation}
V(g\vec{S},g\vec{S}')=V(\vec{S},\vec{S}')=V(\vec{S}',\vec{S}),~~~{\rm
for~all}~g\in SO(n).
\end{equation}
These properties can be used to rewrite the Hamiltonian (7) as
$H_{\rm
spin}\Bigl(\{\vec{S}\}\Bigr)=\sum_{j=1}^{N}V(\vec{S}_{0},g_{j}\vec{S}_{j+1})$
where
$\vec{S}_{0}$ is an arbitrary but fixed unit vector and the $n\times n$
matrices
$g_{j}$ are defined by $g_{j}\vec{S}_{j}:=\vec{S}_{0}$.
With the rotational invariance (10) the partition function (9) can be reduced
to a single $\d\vec{S}$-integration according to
\begin{equation}
Z(\beta )=\int\d\vec{S}_{1}\int\d\vec{S}_{2}\,
\e^{-\beta V(\vec{S}_{0},\vec{S}_{2})}
\cdots
\int\d\vec{S}_{N+1}\,
\e^{-\beta V(\vec{S}_{0},\vec{S}_{N+1})}=\lambda^{N} (\beta )
\end{equation}
where
\begin{equation}
\lambda (\beta ):=
\int\d\vec{S}\,\exp\{-\beta V(\vec{S}_{0},\vec{S})\}.
\end{equation}
Hence, the specific free energy (8) is given by
\begin{equation}
F(\beta )=-(1/\beta )\ln \lambda (\beta ).
\end{equation}

What remains to be done is the integration (14). Choosing as the fixed
vector $\vec{S}_{0}$ the unit vector pointing towards the northpole,
$\vec{S}_{0}=(0,\ldots,0,1)$, the function $V(\vec{S}_{0},\vec{S})$, and
therefore also $\exp\{-\beta V(\vec{S}_{0},\vec{S})\}$, depends
only on the polar angle $\theta $ if $\vec{S}$ is parameterized in the usual
(hyper-) spherical polar coordinates, because then
$\vec{S_{0}}\cdot\vec{S}=\cos\theta $.
Hence, the expression (14) can immediately be reduced to the following
one-dimensional integral ($t:=\cos\theta )$:
\begin{equation}
\lambda(\beta )=\frac{\Gamma (n/2)}{\sqrt{\pi }\,\Gamma (\frac{n-1}{2})}
\int\limits_{-1}^{+1}\d t\,\e^{\beta (Jt+K t^{2})}(1-t^{2})
^{\frac{n-3}{2}}.
\end{equation}
For $K=0$ this integral can be expressed in terms of modified Bessel functions
\begin{equation}
\lambda (\beta )=\Gamma (n/2)\left(\frac{2}{\beta J}\right)^{\frac{n-2}{2}}
\I_{\frac{n-2}{2}}(\beta J),
\end{equation}
the well-known result for Stanley's $n$-vector chain \cite{S69}. For
$K\neq 0$ an expansion in powers of $\beta J$ allows for an integration in
terms of the confluent hypergeometric function \cite{MOS66}:
\begin{equation}
\lambda(\beta )=\sum^{\infty }_{r=0}
\frac{\Gamma (n/2)\,(\beta J/2)^{2r}}{\Gamma (n/2 + r)\,\Gamma(r+1)}
\textstyle\,_{1}F_{1}(r+\frac{1}{2};r+\frac{n}{2};\beta K).
\end{equation}
This series can be summed \cite{Summe} in terms of a generalized
hypergeometric function of two variables
\cite{PBM86,EMOT53}:
\begin{equation}
\textstyle
\lambda (\beta )=\exp\left\{-\frac{\beta J^2}{4K}\right\}\Psi _{2}
\left(\frac{1}{2};\frac{n}{2},\frac{1}{2};\beta K,\frac{\beta J^2}{4K}\right).
\end{equation}
Unfortunately, not much is known about this generalized hypergeometric
function. However, for the cases $n=1,2$ and 3, we can express
$\lambda (\beta )$ somewhat more explicitly in terms of hyperbolic,
modified Bessel and confluent hypergeometric functions, respectively:
\begin{equation}
\begin{array}{lll}
\lambda(\beta )=&\e^{\beta K}\cosh(\beta J)~,&{\rm for}~n=1,\\[4mm]
\lambda(\beta )=&\frac{\e^{\beta K}}{\sqrt{\pi }}
{\displaystyle\sum_{r=0}^{\infty }}
\frac{\Gamma \left(r+\frac{1}{2}\right)}{\Gamma (r+1)}\,
\left(-\frac{2K}{|J|}\right)^r
\I_{r}\left(\beta |J|\right),&{\rm for}~n=2,\\[4mm]
\lambda(\beta )=&\frac{1}{2}\e^{-\beta J^2/4K}\left[
\left(1+\frac{J}{2K}\right)\,
_{1}F_{1}\left(\frac{1}{2};\frac{3}{2};\beta
K(1+\frac{J}{2K})^2\right)\right.&\\[2mm]
&+\left.\left(1-\frac{J}{2K}\right)\,
\,_{1}F_{1}\left(\frac{1}{2};\frac{3}{2};\beta
K(1-\frac{J}{2K})^2\right)\right]
{}~,~&{\rm for}~n=3~.
\end{array}
\end{equation}
For $n=1$ we have used formula 7.2.4.91 of Ref.\ \cite{PBM86} leading to the
expected result. For $n=2$, in essence, we have expanded the integral (16) in
powers of $\beta K$.
For $n=3$ the integral (16) obviously is reducible to the sum of two error
functions with complex argument, which in turn are expressable in terms of
confluent hypergeometric functions. Eqs.\ (19) and (20) in combination with
(15)
summarize the main results we wish to report here.
Appropriate derivatives of $\lambda (\beta )$
with respect to $\beta $ lead to the basic thermostatic quantities. For
example, the specific heat $c(\beta )$ is given as
\begin{equation}
c(\beta )=
-k\beta^{2}\frac{\partial^{2}}{\partial\beta^{2}}
\Bigl(\beta F(\beta )\Bigr)=
k\beta ^{2}\left[\frac{\lambda ''(\beta )}{\lambda (\beta )}-
\left(\frac{\lambda '(\beta )}{\lambda (\beta )}\right)^{2}\right].
\end{equation}
Figure 1 displays this function for $n=3$ as a surface over the
$1/\beta |J|$-$K/|J|$-plane.
We note that its zero-temperature value $c(\infty )=k$ is in agreement
with the classical equipartition theorem. For low temperatures and $K>0$ the
specific heat $c(\beta )$ increases linearly:
\begin{equation}
\frac{c(\beta )}{k}=1+\frac{K}{(K+|J|/2)^{2}\beta }+\cdots .
\end{equation}
For high temperatures it vanishes as the inverse square of the temperature:
\begin{equation}
\frac{c(\beta)}{k}=\left(J^{2}+\frac{4}{15}\,K^{2}\right)
\frac{\beta ^2}{3}+\cdots .
\end{equation}
For $0<K/|J|\,\,\lesssim\,\, 14$ the specific heat attains a maximum value in
the temperature range $0<1/\beta |J|\,\,\lesssim\,\, 0.9$.
This maximum splits for $K/|J|\,\,\gtrsim\,\, 15$ into two maxima.
The global maximum of $c(\beta )$ remains near $1/\beta |J|\,\,\lesssim\,\,1$
for large $K$ values. ~\\[3mm]

Finally, we remark that from the free energy (15) in combination with (16)
one can obtain \cite{JLZ94} further interesting properties of the system (6).
Examples are the two-spin correlation function
\begin{equation}
\langle \vec{S}_{i}\cdot\vec{S}_{j}\rangle=\left(-\frac{\partial}{\partial
J}\,F(\beta )\right)^{|i-j|},
\end{equation}
here $\langle \cdot\rangle$ denotes the canonical equilibrium expectation value
with respect to Hamiltonian (6), the zero-field susceptibility
\begin{equation}
\chi _{0}(\beta ):=\beta \left(1+2\sum_{r=1}^{\infty }
\langle \vec{S}_{j}\cdot\vec{S}_{j+r}\rangle\right)=
\beta\, \frac{1-\frac{\partial}{\partial J}\,F(\beta )}
{1+\frac{\partial}{\partial J}\,F(\beta )},
\end{equation}
and the mean lattice constant
\begin{equation}
a(\beta ):=\ell +\langle q_{j+1}-q_{j}\rangle=\ell -\frac{\eta }{m\omega
_{0}^{2}}\,\frac{\partial}{\partial J}\,F(\beta ).
\end{equation}
A more detailed discussion (including applications and further thermostatic
properties) of the system characterized by the Hamiltonian (6)
will be given elsewhere \cite{JLZ94}.

\clearpage
{\large\bf Figure Caption}
\begin{figure}[h]
\caption{The specific heat (21) for $n=3$ as a function of the dimensionless
temperature $kT/|J|:=1/\beta |J|$ and the parameter $K/|J|$.}
\end{figure}

\end{document}